\documentstyle{elsart}
\newcommand{\rhoL}{\mathrel{\rho^{\mathrm{A}}}}
\newcommand{\rhoR}{\mathrel{\rho^{\mathrm{B}}}}
\input epsf
\begin{document}
\begin{frontmatter}
\title{Front formation in a ballistic annihilation model.}  
\author{Jaros\l aw Piasecki\thanksref{THANK_JAREK}}
\address{Institute of Theoretical Physics, Warsaw University, Ho\.za 69, 
P-00 681 Warsaw, Poland}
\author{Pierre-Antoine Rey and Michel Droz\thanksref{MD}}
\address{Departement de Physique Th\'eorique, Universit\'e de Gen\`eve,
CH-1211 Geneva 4, Switzerland}
\vspace{0.5cm}
\author{UGVA-DPT 1996/01-910}

\thanks[THANK_JAREK]{Work partially supported by the Committee for Scientific Research (Poland), grant 2 P302 074 07.}
\thanks[MD]{Works partially supported by the Swiss National Science 
Foundation}

\begin{abstract}
We study a simple one-dimensional model of ballisticaly-controlled 
annihilation in which the two annihilating species are initially 
spatially separated.  The time dependent properties of the annihilation 
front are exactly derived. It is shown that the front wanders in a 
brownian fashion around its average value.
\end{abstract}


\end{frontmatter}

\section{Introduction}

The study of the kinetics of diffusion-controlled or 
ballisticaly-controlled
annihilation provides a nice playground on which the tools of 
nonequilibrium statistical mechanics can be tested ~\cite{redner,born}. 
Particularly in low dimensions, the kinetics of these systems is 
governed by the fluctuations and any mean-field like (or Boltzmann-like) 
approximation will not give a satisfactory description of the properties 
of the system.

A lot of efforts have be expended in studying the diffusion-controlled 
case, both for the situation in which the reactants are uniformely 
distributed in space~\cite{wil} and
for the one in which they are initially spatially separated, leading to 
reaction-diffusion fronts~\cite{racz,cordro}. In both cases, the long 
time behavior of physical observables is characterized by power law. The 
associated exponents have some universal properties~\cite{born}.

More recently, the case of ballisticaly-controlled annihilation has been 
investigated for spatially homogeneous conditions~\cite{reddue}. In 
particular, exact results have been obtained for one-dimensional, single 
species models by Piasecki~\cite{pia_uno} and Droz, Rey,  Frachebourg 
and Piasecki~\cite{pia_due,pia_tre}.

In these models, one considers point particles which move freely, with a
given velocity.  When two particles collide they instantaneously 
annihilate 
each other and disappear from the system. The system with only two 
possible
velocities $+c$ or $-c$ has been studied in a pioneering work by Elskens 
and
Frisch~\cite{ef}. The case of an arbitrary discrete velocity 
distributions has been investigated by  Droz, Rey,  Frachebourg and 
Piasecki~\cite{pia_tre}. An exact equation for  the survival probability 
until the time $t$ of a particle moving with velocity $v$ was derived, 
allowing to compute the density and the time dependent velocity 
distribution in the asymptotic regime $t\to\infty$.
For a symmetric three velocities distribution,
different kinetic regimes were found as a function of the initial 
fraction of particles at rest. 
Such processes can model several physical situations as  a recombination 
reaction in the gas 
phase or the fluorescence of laser excited gas atoms
with quenching on contact (the one-dimensional aspect can be obtained 
by working in a suitable porous media~\cite{kopel}) or, 
the annihilation of kink-antikink pairs in solid state 
physics\cite{buti}.

A related but different class of problems is  the case in which the two 
species (called $A$ and $B$), are initially separated in space and 
annihilate on contact. Such process can model the situation in which 
chemical species incorporated in a gel move ballistically under the 
action of a drift~\cite{zrini}. As the two species cannot penetrate one 
into the other (as they annihilate on contact), a well defined reaction 
front is formed.
We aim at computing exactly the time dependent properties of this front.

The paper is organized as follows. In section two, the model is defined.
In section 3, we compute the average number of collisions that have 
occured before time $t$. An exact analytic expression for the 
probability density to find the front at a given point and a given time 
is derived.
In section 4, it is shown that the front makes a random walk around its 
(time dependent) average value.
Possible extensions of this work are discussed in the conclusion.

\section{The model}

We consider a one-dimensional system formed of two species of particles.
Initially, particles $A$ are spatially randomly  distributed in the  
region $(-\infty, 0)$
and the $B$ ones are spatially randomly distributed in the  region $(0, 
\infty)$. The positions of the particles obey a Poissonian distribution. 
The
initial linear average densities for the $A$ and $B$ particles are 
respectively $\rhoL$ and $\rhoR$.
The velocities of each  particle is an independent random variable 
taking the value $\pm c$ with even probability. Particles of the same 
kind suffer elastic collisions. When two particles $A$ and $B$ meet, 
they annihilate.
Thus, practically the $A$ particles with velocity $-c$ and the $B$ 
particles with velocity $+c$ move freely. Accordingly, the relevant part 
of the dynamics concerns the $A$ particles with velocity $+c$ and the 
$B$ ones with velocity $-c$.

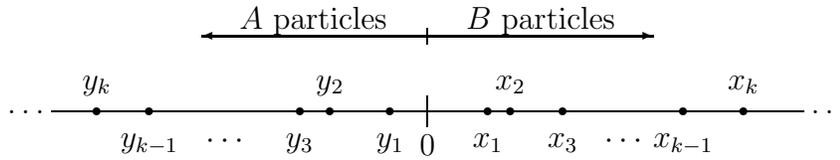
\begin{figure}[tbh]
\setlength{\unitlength}{1cm}
\centerline{
\begin{picture}(12,3.5)(-6,-1)
\put(-5,0){\line(1,0){10}}
\put(0,1){\vector(-1,0){3}}
\put(-2.5,1.1){$A$ particles}
\put(0,1){\vector(1,0){3}}
\put(2.5,1.1){\makebox(0,0)[br]{$B$ particles}}
\put(0,0.9){\line(0,1){0.2}}
\put(0,-0.2){\line(0,1){0.4}}
\put(0,-0.3){\makebox(0,0)[t]{$0$}}
\put(-0.5,0){\circle*{0.1}}
\put(-0.5,-0.3){\makebox(0,0)[t]{$y_1$}}
\put(-1.3,0){\circle*{0.1}}
\put(-1.3,0.3){\makebox(0,0)[b]{$y_2$}}
\put(-1.7,0){\circle*{0.1}}
\put(-1.7,-0.3){\makebox(0,0)[t]{$y_3$}}
\put(-2.7,-0.3){\makebox(0,0)[t]{$\cdots$}}
\put(-3.7,0){\circle*{0.1}}
\put(-3.7,-0.3){\makebox(0,0)[t]{$y_{k-1}$}}
\put(-4.4,0){\circle*{0.1}}
\put(-4.4,0.3){\makebox(0,0)[b]{$y_k$}}
\put(-5.1,0){\makebox(0,0)[r]{$\ldots$}}
\put(0.8,0){\circle*{0.1}}
\put(0.8,-0.3){\makebox(0,0)[t]{$x_1$}}
\put(1.1,0){\circle*{0.1}}
\put(1.1,0.3){\makebox(0,0)[b]{$x_2$}}
\put(1.8,0){\circle*{0.1}}
\put(1.8,-0.3){\makebox(0,0)[t]{$x_3$}}
\put(2.6,-0.3){\makebox(0,0)[t]{$\cdots$}}
\put(3.4,0){\circle*{0.1}}
\put(3.4,-0.3){\makebox(0,0)[t]{$x_{k-1}$}}
\put(4.2,0){\circle*{0.1}}
\put(4.2,0.3){\makebox(0,0)[b]{$x_k$}}
\put(5.1,0){\makebox(0,0)[l]{$\ldots$}}
\end{picture}}
\caption{Typical initial condition. The particles are labeled such that 
$y_k<\cdots<y_1<0<x_1<\cdots<x_k$. \label{fig:one}}
\end{figure}

Let $(y_1,y_2,\ldots,y_k,\ldots)$ be the initial positions of the 
$A$ particles with velocity $+c$ and 
$(x_1,x_2,\ldots,x_k,\ldots)$ the initial positions of the $B$'s with 
velocity $-c$ (see figure~\ref{fig:one}). The relative velocity between 
the $A$ and 
$B$ particles being $2c$, the pair of particles initially at $(x_k, 
y_k)$ will collide at time $t_k=(x_k-y_k)/2$.

The position at which this collision will take place  defines the 
position of the annihilation front.
Thus, the position of the 
front at time $t$ is $\half(x_1-y_1)$, if $2ct\leq 
x_1+y_1$ (i.e. if no particle have yet collided), $\half(x_2-y_2)$, if 
$x_1+y_1\leq 2ct\leq x_2+y_2$ (i.e.  only 
the first particles on the right and on the left have collided), and so 
on. 
The dynamics in itself is purely deterministic, the only stochastic 
aspect comes from the initial conditions.
The properties of the front are completely defined by the  
probability density $\mu(X;t)$ to find the front at the point $X$, 
at time $t$. Moreover, another quantity of interest is  $N(t)$, the 
number of  particles that have been annihilated until a given time $t$.

\section{Analytical solution}\label{sec:method}

We first investigate the behavior of $N(t)$.
For this purpose, we introduce   the probability $P(k;t)$
that exactly $k$ collisions have occured within the time interval 
$[0,t]$. For a given initial configuration, at least $k$ collisions have 
occured before the time $t$, if $2ct\geq x_k-y_k$, and  at most $k$ 
collisions have occured before $t$, if $2ct\leq x_{k+1}-y_{k+1}$, 
with $x_k<x_{k+1}$ and $y_k>y_{k+1}$. Hence averaging over the initial 
conditions, we find:
\begin{eqnarray}
P(k;t) &=& \langle\theta\left(2ct-[x_k-y_k]\right)
                  \theta\left([x_{k+1}-y_{k+1}]-2ct\right)\rangle        
           \nonumber\\
       &=& \langle\theta\left(2ct-[x_k-y_k]\right)\rangle -
           \langle\theta\left(2ct-[x_{k+1}-y_{k+1}]\right)\rangle\, .
           \label{eq:kcoll}
\end{eqnarray}
where $\theta$ is the usual Heaviside function and the brackets denote 
the average over the initial conditions (see Appendix~\ref{app:B}).
Thus:
\begin{equation}
N(t)=\sum_{k=1}^\infty k P(k;t)\, ,
\end{equation}
which, using (\ref{eq:kcoll}), reduces to \begin{equation}
N(t)=\sum_{k=1}^\infty \langle\theta\left(2ct-[x_k-y_k]\right)\rangle
\, .
\end{equation}
Thus, using the results of Appendix~\ref{app:B}, $N(t)$ reads:
\begin{eqnarray*}
N(t)=\sum_{k=1}^\infty \int_{-\infty}^0 \d y_k\rhoL \int_0^\infty \d x_k
&\rhoR& 
     \e^{-\rhoL|y_k|}\frac{\left(|y_k|\rhoL\right)^{k-1}}{(k-1)!} \\
&\times&
     \e^{-\rhoR x_k}\frac{\left(x_k\rhoR\right)^{k-1}}{(k-1)!}
     \theta\left(2ct-[x_k-y_k]\right)\, .
\end{eqnarray*}
Noticing that the sum over $k$ can be exchanged with the two integrals, 
we 
find (where $a=\half(x_k-y_k)$ and $b=\half(x_k+y_k)$)
\[
N(t)=2\rhoL\rhoR \int_0^{ct} \d a \int_{-a}^a \d b
     \e^{-a(\rhoL+\rhoR)} \e^{b(\rhoL-\rhoR)}
     I_0\left(2\sqrt{\rhoL\rhoR(a^2-b^2)}\right)\, ,
\]
$I_0$ is the modified Bessel function. The integral over $b$ is  
calculated 
in appendix~\ref{app:A} and the integration on $a$ is straightforward. 
We obtain
\begin{equation}
N(t)=\frac{2\rhoL\rhoR}{\rhoL+\rhoR}\left[ct - \frac{1}{2(\rhoL+\rhoR)}
     \left(1-\e^{-2ct(\rhoL+\rhoR)}\right)\right]\, .\label{eq:nbcol}
\end{equation}
In terms of the dimensionless time
\[
\tau\equiv ct(\rhoL+\rhoR)
\]
and the parameter
\[
\lambda\equiv\frac{4\rhoL\rhoR}{\rhoL+\rhoR}\, ,
\]
we have:
\begin{equation}
N(\tau)=\frac{\lambda}{2}\left[\tau -
        \frac{1}{2}\left(1-\e^{-2\tau}\right)\right]\, .
\end{equation}
After a transient regime, $N(\tau)$ grows linearly with time.

The front position is defined as the point where the next collision 
will 
take place. Thus the probability density $\mu(X;t)$ to find the front at 
the point $X$, 
at time $t$ is  given by
\begin{eqnarray}
\lefteqn{\mu(X;t) =
    \left\langle\delta\left(X-\frac{x_1+y_1}{2}\right)
    \theta(x_1-y_1-2ct)\right\rangle}\nonumber \\
&+& \left\langle\sum_{k=1}^\infty \delta\left(X 
-
             \frac{x_{k+1}+y_{k+1}}{2}\right)
    \theta\left(2ct-[x_k-y_k]\right)
    \theta\left([x_{k+1}+y_{k+1}]-2ct\right)\right\rangle \nonumber \\
&=& 
\left\langle\delta\left(X-\frac{x_1+y_1}{2}\right)\right\rangle
    \label{eq:defmu} \\
&+& \left\langle\sum_{k=1}^\infty \left[
             \delta\left(X-\frac{x_{k+1}+y_{k+1}}{2}\right) -
             \delta\left(X-\frac{x_k+y_k}{2}\right) \right]
    \theta\left(2ct-[x_k-y_k]\right)\right\rangle\, . \nonumber
\end{eqnarray}
The first term in the right hand side reads simply
\begin{eqnarray}
\left\langle\delta\left(X - \frac{x_1+y_1}{2}\right)\right\rangle
&=& \rhoL\rhoR \int_0^\infty \d x_1 \int_{-\infty}^0 \d y_1 
       \e^{-\rhoR x_1}\e^{-\rhoL y_1}
       \delta\left(X-\frac{x_1+y_1}{2}\right) \nonumber\\
&=& \frac{2\rhoL\rhoR}{\rhoL+\rhoR}\left[\theta(X)\e^{-2\rhoR X} +
       \theta(-X)\e^{2\rhoL X}\right]\, .
\end{eqnarray}

To calculate the second term in the right hand side, we remark that one 
has to
 average over the position $x_k$, $x_{k+1}$, $y_k$ and $y_{k+1}$. We 
shall then first average over particles $k+1$, to obtain an expression 
containing only the position $x_k$ and $y_k$. To do this, we need to 
calculate the following integral
\[
S \equiv \rhoR\int_{x_k}^\infty \d x_{k+1} \rhoL\int_{-\infty}^{y_k} \d 
y_{k+1}
\delta\left(X-\frac{x_{k+1}+y_{k+1}}{2}\right)
\e^{-\rhoR x_{k+1}+\rhoL y_{k+1}}\, ,
\]
putting $a=\half(x_{k+1}-y_{k+1})$ and 
$b=\half(x_{k+1}+y_{k+1})$, we have:
\begin{eqnarray*}
S = 2\rhoL\rhoR\int_{-\infty}^\infty \d a \int_{-\infty}^\infty \d b 
\theta(a+b-x_k)&\theta&(a-b+y_k)\delta(x-b) \\
&\times&\e^{-(\rhoL+\rhoR)a+(\rhoL-\rhoR)b}
\end{eqnarray*}
The integral over $b$ is straightforward, and the integral over $a$ 
gives after some algebra
\begin{eqnarray*}
S = \frac{2\rhoL\rhoR}{\rhoL+\rhoR}
&& \left[\rule{0mm}{5mm}\theta\left(\frac{x_k+y_k}{2}-X\right)
   \e^{-(\rhoL+\rhoR)x_k+2\rhoL X}\right .\\
&& +\left.\rule{0mm}{5mm}\theta\left(X-\frac{x_k+y_k}{2}\right)
   \e^{(\rhoL+\rhoR)y_k-2\rhoR X}\right]\, .
\end{eqnarray*}
Hence, the second term in the right hand side of eq.~(\ref{eq:defmu}) 
becomes
\begin{eqnarray*}
\sum_{k=1}^\infty\left\langle\left\{\rule{0mm}{6mm}
   \frac{2\rhoL\rhoR}{\rhoL+\rhoR}\right.\right.
&& \left[\rule{0mm}{5mm}\theta\left(\frac{x_k+y_k}{2}-X\right)
   \e^{-(\rhoL+\rhoR)x_k+2\rhoL X}\right .\\
&& 
+\left.\rule{0mm}{5mm}\theta\left(X-\frac{x_k+y_k}{2}\right)
   \e^{(\rhoL+\rhoR)y_k-2\rhoR 
X}\right]\\
&& -\left.\left.\rule{0mm}{6mm}
   \delta\left(X-\frac{x_k+y_k}{2}\right) \right\}
    \theta\left(2ct-[x_k-y_k]\right)\right\rangle\, .
\end{eqnarray*}
Now using the relation (see appendix~\ref{app:B})
\begin{eqnarray}
\lefteqn{\sum_{k=1}^\infty\langle f(x_k+y_k,x_k-y_k) \rangle}
    \label{eq:sumav} \\
&=& 2\rhoL\rhoR\int_0^\infty \d a \int_{-a}^a \d b
    e^{-(\rhoL+\rhoR)a+(\rhoL-\rhoR)b} f(2b,2a)
    I_0\left(2\sqrt{\rhoL\rhoR(a^2-b^2)}\right)\, ,\nonumber
\end{eqnarray}
one finally finds:
\begin{eqnarray}
\lefteqn{\mu(X;t) = \frac{2\rhoL\rhoR}{\rhoL+\rhoR}\left[\theta(X)
         \e^{-2\rhoR X}+\theta(-X)\e^{2\rhoL X}\right]} \\
     &+& 2\rhoL\rhoR \int_0^{ct} \d a \int_{-a}^a \d b
         \e^{-(\rhoL+\rhoR)a+(\rhoL-\rhoR)b}
         I_0\left(2\sqrt{\rhoL\rhoR(a^2-b^2)}\right)
         \nonumber \\
&\times& \left\{-\delta(X-b)+\frac{2\rhoL\rhoR}{\rhoL+\rhoR}
         \left[\theta(b-X)e^{-2\rhoL(b-X)}+
         \theta(X-b)e^{-2\rhoR(X-b)}\right]\right\}\, . \nonumber
\end{eqnarray}
In terms of the dimensionless  variables $\tau$ and $\lambda$, and 
 the dimensionless coordinate $\xi\equiv(\rhoL+\rhoR)X$, we can 
write
\begin{eqnarray}
\mu(\xi;\tau)
&=& \frac{\rhoL+\rhoR}{2}\e^{\kappa\xi} \left\{\rule{0mm}{6mm}
    \lambda\e^{-|\xi|}\right. \label{eq:mugen} \\
&+& \left.\rule{0mm}{6mm}\lambda\int_0^\tau \d\alpha \e^{-\alpha}
     \int_{-\alpha}^\alpha \d\beta
     I_0\left(\sqrt{\lambda(\alpha^2-\beta^2)}\right)
     \left[-\delta(\beta-\xi)+
     \frac{\lambda}{2}\e^{-|\beta+\xi|}\right]\right\}\, ,\nonumber
\end{eqnarray}
with
\[
\kappa\equiv\frac{\rhoL-\rhoR}{\rhoL+\rhoR}\, .
\]
Note that the function between 
the curly brackets is an even function of $\xi$; we shall call it 
$\hat{\mu}(\xi;\tau)$. The rest of the paper will be devoted to the 
study of this function.

\section{Results and interpretation}\label{sec:results}

Starting from~(\ref{eq:mugen}) and using the integrals quoted in  
appendix~\ref{app:A}, it is possible to verify that $\mu(x;t)$ is 
correctly normalized to $1$. Moreover, we can also calculate its 
different 
moments. However for this purpose, it is simpler to consider the Fourier 
transform of $\mu$.
Let us define the function ${\tilde{\mu}}(p;\tau)$ as:
\begin{eqnarray}
{\tilde{\mu}}(p;\tau) &=& \int_{-\infty}^\infty \d X 
\e^{-(\rhoL+\rhoR)pX}
                \e^{-(\rhoL-\rhoR)X} \mu(X;\tau) \nonumber \\
            &=& \frac{1}{2}\int_{-\infty}^\infty \d\xi \e^{-p\xi}
                \hat{\mu}(\xi;\tau)\, .\label{eq:defmufou}
\end{eqnarray}
Here again, this integral can be calculated using the formulas quoted in
appendix~\ref{app:A}. One finds:
\begin{equation}
{\tilde{\mu}}(p;\tau)=\lambda\e^{-\tau}\frac{1}{1-p^2}\left[
            \cosh\left(\sqrt{p^2+\lambda}\tau\right)+
            \frac{\sinh\left(\sqrt{p^2+\lambda}\tau\right)}
                 {\sqrt{p^2+\lambda}}\right]\label{eq:mufou}
\end{equation}
It is now easy  to calculate the different moment of $\mu$, using the 
fact that
\begin{equation}
\langle\xi^n\rangle
  \equiv \int_{-\infty}^\infty \d X\left[(\rhoL+\rhoR)X\right]^n\mu(X;t)
  =(-1)^n\left[\frac{\partial^n}{\partial p^n}{\tilde{\mu}}(p;\tau)
  \right]_{p=-\kappa}\, .\label{eq:defmom}
\end{equation}
Noticing that  $\kappa^2+\lambda=1$, it is straightforward to check the 
normalization. The first moment ${\bar{\xi}}(\tau)=\langle\xi\rangle$ is 
given by
\begin{equation}
{\bar{\xi}}(\tau) = \kappa\left[\frac{2}{\lambda}+\tau-\frac{1}{2}
            \left(1-\e^{-2\tau}\right)\right]
          = \frac{2\kappa}{\lambda}\left[1+N(\tau)\right]\, ,
\end{equation}
and the variance is
\begin{eqnarray}
\lefteqn{\xi_{\mathrm{rms}}^2(\tau) = \langle\xi^2\rangle - 
\langle\xi\rangle^2} \\
&=& \frac{4}{\lambda^2} - \frac{2}{\lambda} + \frac{3}{4} -
    \frac{5\lambda}{4} 
+\lambda\tau 
+
    \left[\lambda-\frac{1}{2}-2\tau(1-\lambda)\right]\e^{-2\tau} -
    \frac{1-\lambda}{4}\e^{-4\tau}\, . \nonumber
\end{eqnarray}
If $\rhoR \not= \rhoL$, and   in the long time limit, the front moves 
with a velocity proportionnal to the density difference.
The case $\lambda=1$ is of particular interest, because it corresponds 
to the symetric case  $\rhoL=\rhoR$. As expected: 
\begin{equation}
{\bar{\xi}}(\tau)\equiv 0
\end{equation}
but
\begin{equation}
\xi^2_{\mathrm{rms}}(\tau) = 2\left[1+N(\tau)\right]\, .
\end{equation}
In the long time limit, we see that 
$\xi_{\mathrm{rms}}^2(\tau)\propto\tau$:
the front move essentially as an unbiased random walker. Thus, we can 
expect
 $\mu$ to be essentially a gaussian. In fact, we shall prove that
for any $\lambda$, the long time limit of $\mu$ is a gaussian.  
Starting from the definition of an arbitrary moment of $\mu$ 
(eq.~(\ref{eq:defmom})), we may write for long time:
\begin{equation}
\langle\xi^n\rangle=(-1)^n\lambda\e^{-\tau}\left[\frac{\partial^n}
{\partial p^n}\e^{\tau\sqrt{p^2+\lambda}}\right]_{p=-\kappa}
+ O(\tau^{n-1})\, ,
\end{equation}
using
\begin{equation}
\frac{\partial^n}{\partial p^n}\e^{\tau\sqrt{p^2+\lambda}}=
\left[\frac{\partial}{\partial p}\sqrt{p^2+\lambda}\right]^n
\left[\frac{\partial^n}{\partial \eta^n}\e^{\tau\eta}
\right]_{\eta=\sqrt{p^2+\lambda}} + O(\tau^{n-1})\, ,
\end{equation}
we obtain $\langle\xi^n\rangle=(\kappa\tau)^n+O(\tau^{n-1})$.

If $\lambda<1$, $\kappa\neq 0$ and we know the leading term for every 
moment of $\mu$. If $\lambda=1$, the situation is different and will be 
considered later.

It is known that for a gaussian distribution with a non vanishing mean 
${\bar{\xi}}(\tau)$ 
and a variance $\xi_{\mathrm{rms}}^2(\tau)$, one has:
\begin{equation}
\langle\xi^n\rangle_{\mathrm{gauss}}=n!\xi^n(\tau)
\sum_{k=0}^{[\frac{n}{2}]}\frac{1}{(n-2k)!k!}
\left(\frac{\xi_{\mathrm{rms}}^2(\tau)}{2\xi^2(\tau)}\right)^k\, ,
\end{equation}
In the long time limit:
\begin{equation}
\frac{\xi_{\mathrm{rms}}^2(\tau)}{2{\bar{\xi}}^2(\tau)}
=O\left(\frac{1}{\tau}\right)\, ,
\end{equation}
and $\langle\xi^n\rangle_{\mathrm{gauss}}=(\kappa\tau)^n+O(\tau^{n-1})$.

Thus, when $\lambda <1$  and for sufficiently long times (i.e. when the 
corrections to the leading 
term become negligible), the probabilty density $\mu$ takes the gaussian 
form:
\begin{equation}
\mu(\xi;\tau)\propto
\exp\left[-\frac{(\xi-\kappa\tau)^2}{2\lambda\tau}\right]\, .
\end{equation}
The front wanders around its mean value with an amplitude of order of
magnitude $\sqrt{\lambda\tau}$, like a biased random walker  with a 
diffusion coefficient of $\lambda/2$.

The $\lambda=1$ case is different because the amplitude of the $\tau^n$ 
term vanishes. However, we can still prove that for large time, the 
probability density $\mu$ is a gaussian. More precisely, we shall give 
an upper and a lower bound, both of gaussian form. For this purpose, we 
shall come back to the Fourier transform ${\tilde{\mu}}(p;\tau)$ defined 
by 
eq.~(\ref{eq:defmufou}). It can be shown (see appendix~\ref{app:C}), 
that the inverse Fourier transform of ${\tilde{\mu}}(p,\tau)$ is given 
by
\begin{eqnarray}
\hat\mu(\xi;\tau) &=&\lambda\e^{-|\xi|-\tau}
                     \left[\cosh\left(\sqrt{1+\lambda}\tau\right) +
                     \frac{\sinh\left(\sqrt{1+\lambda}\tau\right)}
                     {\sqrt{1+\lambda}}\right] \label{eq:muborne}\\
                  &-&\lambda\theta(\tau-|\xi|)\e^{-\tau}
                     \left\{\rule{0mm}{6mm}
                     \sinh\left(\sqrt{1+\lambda}\tau-|\xi|\right) +
                     \frac{\cosh\left(\sqrt{1+\lambda}\tau-|\xi|\right)}
                     {\sqrt{1+\lambda}}\right.\nonumber \\
                  &-&\left.\rule{0mm}{6mm}
                     2\int_0^{\sqrt{\lambda}}\frac{\d r}{\pi}
                     \frac{\cos(r\xi)}{1+r^2}\left[
                     \sinh\left(\sqrt{\lambda-r^2}\tau\right)+
                     \frac{\cosh\left(\sqrt{\lambda-r^2}\tau\right)}
                     {\sqrt{\lambda-r^2}}\right]\right\}\, .\nonumber
\end{eqnarray}
It is now easy to find an upper and a lower bound for $\hat\mu$. Indeed, 
 for $|\xi|\leq\frac{\pi}{2\sqrt{\lambda}}$, one has:
\[
\frac{\cos(r\xi)}{1+\lambda}\leq\frac{\cos(r\xi)}{1+r^2}\leq\cos(r\xi)
\]
as $r\in[0,\sqrt{\lambda}]$.
The remaining integral can be calculated exactly:
\begin{eqnarray}
\lefteqn{\frac{2}{\pi}\int_0^{\sqrt{\lambda}} \d r 
\cos(r\xi)\left[
         \sinh\left(\sqrt{\lambda-r^2}\tau\right)+
         \frac{\cosh\left(\sqrt{\lambda-r^2}\tau\right)}
              {\sqrt{\lambda-r^2}}\right]}\nonumber \\
\quad
&=& \lambda\tau\frac{I_1\left(\sqrt{\lambda(\tau^2-\xi^2)}\right)}
                    {\sqrt{\lambda(\tau^2-\xi^2)}}+
    I_0\left(\sqrt{\lambda(\tau^2-\xi^2)}\right)\, , \label{eq:borne}
\end{eqnarray}
yielding, then two bounds for $\mu$. Two conclusions can then be
drawn. First for $\lambda=1$, the asymptotic time behavior 
of~(\ref{eq:borne}) is
\[
\e^\tau\frac{2}{\sqrt{2\pi\tau}}\e^{-\xi^2/2\tau}\, ,
\]
and then, for $\xi\leq\frac{\pi}{2}$ and $\tau$ large, we have:
\begin{equation}
\frac{1}{\sqrt{2\pi\tau}}\e^{-\xi^2/2\tau}\leq\hat\mu(\xi;\tau)
\leq\frac{2}{\sqrt{2\pi\tau}}\e^{-\xi^2/2\tau}\, .
\end{equation}
In the long time limit, the density probability $\mu$ becomes the 
density 
probability of a random walk, with a diffusion coefficient of $\half$.

The second consequence concerns the case $\xi=0$, for any value of 
$\lambda$: we have, in the long time limit:
\begin{eqnarray}
\hat\mu(0;\tau) &\propto& 
\frac{\e^{(1-\sqrt{\lambda})\tau}}{\sqrt{\tau}}
                          \quad\mbox{if }\lambda<1\, , \\
\hat\mu(0;\tau) &\propto& \frac{1}{\sqrt{\tau}}
                          \quad\mbox{if }\lambda=1\, .
\end{eqnarray}
These results are not really surprising. Indeed, it can be verified that 
$\xi=0$ is 
a maximum of $\mu$, for $\lambda=1$ but not for $\lambda<1$.

\begin{figure}
\epsfysize=8cm
\centerline{\epsfbox{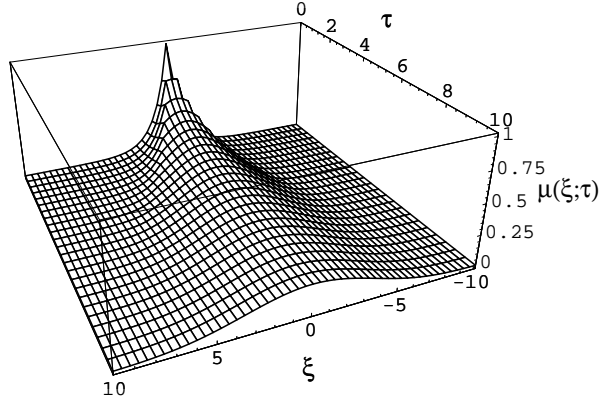}}
\caption{Plot of $\mu(\xi;\tau)$, for $\rhoL=1$ and $\rhoR=1$
(i.e. $\lambda=1$). 
\label{fig:two}}
\end{figure}

\begin{figure}
\epsfysize=8cm
\centerline{\epsfbox{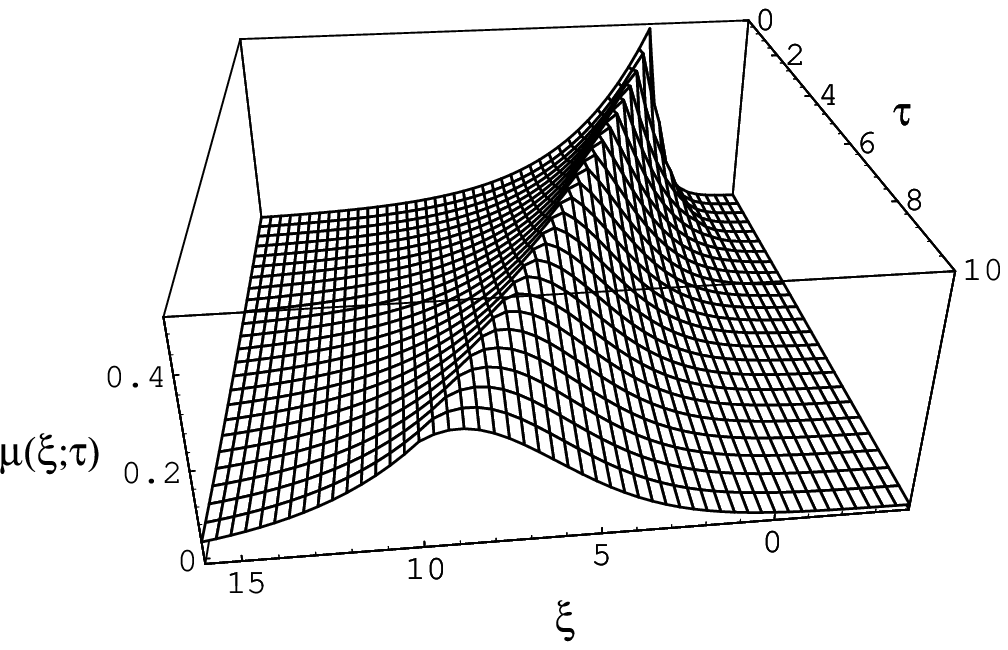}}
\caption{Plot of $\mu(\xi;\tau)$, for $\rhoL=1+\half\protect\sqrt{3}$ and 
$\rhoR=1-\half\protect\sqrt{3}$ (i.e. $\lambda=0.5$). 
\label{fig:three}}
\end{figure}

It is useful to get an image of what is going on by plotting the 
function
$\mu$ for various value of $\lambda$. However for this purpose, both 
equations~(\ref{eq:mugen}) and~(\ref{eq:muborne}) are inadequate for 
technical reasons. Equation~(\ref{eq:mugen}) is not well suited
 because it contains a double integral which requires a lot of 
CPU time and equation ~(\ref{eq:muborne})  because the integral diverge 
for $r=\sqrt\lambda$, 
leading to numerical difficulties.
To avoid these problems, we  have  obtained a different 
integral representation, starting from~(\ref{eq:mugen}), and using the 
following relation:
\begin{equation}
\left[\frac{\partial^2}{\partial\alpha^2} -
      \frac{\partial^2}{\partial\beta^2}\right]
I_0\left(\sqrt{\lambda(\tau^2-\xi^2)}\right) =
\lambda I_0\left(\sqrt{\lambda(\tau^2-\xi^2)}\right)\label{eq:Ider} \, .
\end{equation}
This relation follows directly from the integral representation for 
$I_0$
\[
I_0\left(\sqrt{\lambda(\tau^2-\xi^2)}\right) =
\int_{\mathrm{C}} \frac{\d z}{2\mathrm{i}\pi}\frac{1}{z}
\exp\left\{\frac{\sqrt{\lambda}}{2}\left[(\alpha+\beta)z+
           \frac{\alpha-\beta}{z}\right]\right\}\, ,
\]
where $\mathrm{C}$ is, for example, the circle of radius $1$ centered
in $z=0$. Inserting~(\ref{eq:Ider}) in~(\ref{eq:mugen}), integrating 
twice by part and using that
\[
\left[\frac{\partial^2}{\partial\alpha^2} - 
      \frac{\partial^2}{\partial\beta^2}\right]
\e^{-\alpha-|\beta-\xi|} = 2\e^{-\alpha}\delta(\beta-\xi)\, ,
\]
we get
\begin{eqnarray*}
\hat\mu(\xi;\tau) = \lambda\left\{\e^{-|\xi|}\rule{0mm}{10mm}\right.
    + \frac{1}{2}\int_0^\tau
&\d&\alpha
    \left[\e^{-\tau}\left(\e^{-|\alpha-\xi|}+\e^{-|\alpha+\xi|}\right)
    \rule{0mm}{9mm}\right. \\
&\times&
    \left(\lambda\tau
    \frac{I_1\left(\sqrt{\lambda(\tau^2-\alpha^2)}\right)}
         {\sqrt{\lambda(\tau^2-\alpha^2)}}+
    I_0\left(\sqrt{\lambda(\tau^2-\alpha^2)}\right)\right)\\
&-& \left.\rule{0mm}{10mm}\left.\rule{0mm}{9mm}
    2\theta(\alpha-\xi)e^{\xi-2\alpha} 
-
    2\theta(\alpha+\xi)e^{2\alpha-\xi}\right]\right\}\, .
\end{eqnarray*}
Performing the integral over the two theta functions, rearranging 
the terms in the remaining integral and using the integrals of 
appendix~\ref{app:A}, leads to:
\begin{eqnarray}
\hat\mu(\xi;\tau)
&=& \lambda\e^{-|\xi|-\tau}
    \left[\cosh\left(\sqrt{1+\lambda}\tau\right) +
    \frac{\sinh\left(\sqrt{1+\lambda}\tau\right)}
          {\sqrt{1+\lambda}}\right] \label{eq:muplot}\\
&-& \lambda\theta(\tau-|\xi|)\e^{-\tau}\left\{\rule{0mm}{10mm}
    \sinh(\tau-\xi)\right.\nonumber \\
&-& \left.\rule{0mm}{10mm}
    \int_{|\xi|}^{\tau} \d\alpha\sinh(\alpha-|\xi|)\left[
    \frac{I_1\left(\sqrt{\lambda(\tau^2-\alpha^2)}\right)}
         {\sqrt{\lambda(\tau^2-\alpha^2)}}+
    I_0\left(\sqrt{\lambda(\tau^2-\alpha^2)}\right)\right]\right\}\, .
    \nonumber
\end{eqnarray}
We have not been able to simplify further this expression and in 
particular to perform analytically the integration. However, this form 
can be easily integrated numerically. For examples, the function 
$\mu(\xi;\tau)$ is plotted on figure~\ref{fig:two} and~\ref{fig:three} 
for respectively
$\lambda=1$ and for $\lambda=0.5$, $\rhoL>\rhoR$.

One sees that for $\lambda=1$, the probability density $\mu$ spreads out 
and approches a gaussian for long times. For the assymetric case 
$\lambda=0.5$, one sees clearly the drift of the front which beecomes 
linear in the long time regime.

\section{Conclusions}\label{sec:concl}

Using a combinatorial analysis approach, we have been able to derive exact
results for the time dependent number of collisions, the probability
density $\mu(X;t)$ to find the front at point $X$ at time $t$ and
its first two moments.
Although simple, this nonequilibrium model leads to nontrivial behavior 
as far as the dynamics of the annihilation front is concerned.  

Several extensions of the present work are possible. One is to consider
a similar model in 
which the velocities are not restricted to be bimodal, but obey  a more 
general (discrete or continuous) distribution. Another interesting 
aspect is given by the generalization of  this model to higher 
dimensions. In this case, some $A$ particles can invade the right part 
of the system whithout being annihilated (and vice-versa for the $B$'s). 
A new definition of the front in term of the annihilation rate should be 
introduced and different time dependent behavior may occur. These more 
complicated problems are under investigation.

\appendix{
\section{Appendix}\label{app:B}

The average over the initial conditions is taken as follows: the $B$ 
particles are distributed to the right of the origin, with a Poissonian 
distribution of density $\rhoR$. Thus the probability to find the first 
particle at position $x_1$ is
\[
\pi_1(x_1, \rhoR)=\e^{-\rhoR x_1}\rhoR\, .
\]
The probability to find $k-1$ particles between $0$ and $x_k$ and a 
particle in $x_k$ is
\[
\pi_k(x_k,\rhoR)=\e^{-\rhoR x_k}\frac{\left(x_k\rhoR\right)^{k-1}}{(k-1)!}\rhoR\, .
\]
The average of a function depending only on $x_k$ is thus
\begin{equation}
\langle f(x_k)\rangle=\int_0^\infty \d x_k f(x_k)
\e^{-\rhoR x_k}\frac{\left(x_k\rhoR\right)^{k-1}}{(k-1)!}\rhoR\, .
\end{equation}
Now, the average of a function depending both on $x_k$ and $x_{k+1}$ is
\begin{equation}
\langle f(x_k,x_{k+1})\rangle=
\int_0^\infty \d x_k \int_{x_k}^\infty \d x_{k+1} f(x_k,x_{k+1})
\e^{-\rhoR x_{k+1}}\frac{\left(x_k\rhoR\right)^{k-1}}{(k-1)!}\rhoR^2\, .
\end{equation}
Indeed, the probability to find exactly $k-1$ particles between $0$ and 
$x_k$, one in $x_k$ and another in $x_{k+1}$, but none between is
\[
\e^{-\rhoR (x_{k+1}-x_k)}\rhoR\quad\e^{-\rhoR 
x_k}\frac{\left(x_k\rhoR\right)^{k-1}}{(k-1)!}\rhoR\, ,
\]
as $\e^{-\rhoR (x_{k+1}-x_k)}\rhoR$ is the probability that the interval 
$[x_k,x_{k+1}]$ is empty.
Similar considerations apply for the $A$ particles.

Now, we propose to prove the formula~(\ref{eq:sumav}).
\begin{eqnarray}
\lefteqn{\sum_{k=1}^\infty\langle f(x_k+y_k,x_k-y_k)\rangle}
         \nonumber\\
&=& \sum_{k=1}^\infty \rhoL\rhoR \int_0^\infty \d x_k 
    \int_{-\infty}^0 \d y_k \e^{-\rhoR x_k+\rhoL y_k}
    \frac{\left[\rhoR x_k\rhoL (-y_k)\right]^{k-1}}
         {\left[(k-1)!\right]^2} \nonumber \\
&& \hspace*{3.8cm}\times f(x_k+y_k,x_k-y_k) \\
&=& 2\rhoL\rhoR \int_{-\infty}^{\infty} \d a 
                \int_{-\infty}^{\infty} \d b \theta(a+b) \theta(a-b)
    \e^{-(\rhoL+\rhoR)a+(\rhoL-\rhoR)b} \nonumber \\
&& \hspace*{3.2cm}\times f(2b,2a)
    I_0\left(2\sqrt{\rhoL\rhoR(a^2-b^2)}\right) \label{eq:Btwo} \\
&=& 2\rhoL\rhoR \int_0^{\infty} \d a \int_{-a}^{a} \d b
    \e^{-(\rhoL+\rhoR)a+(\rhoL-\rhoR)b} f(2b,2a) \nonumber \\
&&  \hspace*{3cm}\times I_0\left(2\sqrt{\rhoL\rhoR(a^2-b^2)}\right)\, ,
\end{eqnarray}
where in~(\ref{eq:Btwo}), we put $2b=x_k+y_k$ and $2a=x_k-y_k$ and use
the serie representation of $I_0$.

\section{Appendix }\label{app:A}

In this appendix, we quote two integrals that we often use in our 
derivation. The first is used, for example, in deriving the formula for 
$N(t)$~(\ref{eq:nbcol}), or in calculating the different moments
of $\mu$ without introducing the Fourier transform.
\begin{eqnarray}
\frac{1}{2}\int_{-y}^y \d x \e^{\beta_1 x}
           I_0\left(\beta_2\sqrt{y^2-x^2}\right)
&=& \int_0^y \d x \cosh(\beta_1 x)
    I_0\left(\beta_2\sqrt{y^2-x^2}\right) \\
&=& \frac{\sinh\left(y\sqrt{\beta_1^2+\beta_2^2}\right)}
         {\sqrt{\beta_1^2+\beta_2^2}}
\end{eqnarray}
(see for example~\cite{rus} equation $8$).

In calculating~(\ref{eq:Ider}), we used the previous integral and also
the following
\begin{equation}
\int_0^{\tau} \d \alpha \cosh(\alpha\kappa)\lambda\tau
\frac{I_1\left(\sqrt{\lambda(\tau^2-\alpha^2)}\right)}
     {\sqrt{\lambda(\tau^2-\alpha^2)}} =
\cosh(\tau)-\cosh(\kappa\tau)\, ,
\end{equation}
which can be found in~\cite{rus} equation $6  $).

\section{Appendix }\label{app:C}

In this appendix, we shall prove formula~(\ref{eq:muborne}), starting 
from eq.~(\ref{eq:mufou}). Remembering the 
definition~(\ref{eq:defmufou}) of ${\tilde{\mu}}(p;\tau)$, we may write
\begin{eqnarray*}
\hat\mu(\xi;\tau) = \lambda\e^{-\tau}
&\int_{-\mathrm{i}\infty-\kappa}^{\mathrm{i}\infty-\kappa}&
    \frac{\d p}{2\mathrm{i}\pi}
    \e^{\xi p}\left[\frac{1}{p+1}-\frac{1}{p-1}\right]\\
&\times&  
    \left[\cosh\left(\sqrt{p^2+\lambda}\tau\right)+
    \frac{\sinh\left(\sqrt{p^2+\lambda}\tau\right)}{\sqrt{p^2+\lambda}}
    \right]\, .
\end{eqnarray*}
The integrant has two poles ($p=\pm 1$) and a cut on the imaginary 
axis between $-\mathrm{i}\sqrt{\lambda}$ and $\mathrm{i}\sqrt{\lambda}$. 
For $\xi>\tau$, we can close the contour to the left, without changing 
the value of the integral. We get then a contribution from the pole 
$p=-1$ and eventually from the cut (depending on the sign of $\kappa$). 
Nevertheless, as $\cosh(x)+\sinh(x)/x$ is an even function, the 
contribution of the cut vanishes and we are left with
\begin{equation}
\hat\mu(\xi;\tau) = 
\lambda\e^{-\tau-\xi}
                    \left[\cosh\left(\sqrt{1+\lambda}\tau\right) +
                    \frac{\sinh\left(\sqrt{1+\lambda}\tau\right)}
                    {\sqrt{1+\lambda}}\right]\, ,\quad\xi>\tau\, .
                    \label{eq:Cone}
\end{equation}
For $\xi\leq\tau$, we close the contour to the right, leading to a 
contribution of the pole $p=+1$ only, as, again, the contribution of the 
cut vanishes. We find
\begin{equation}
\hat\mu(\xi;\tau) = \lambda\e^{-\tau-|\xi|}
                    \left[\cosh\left(\sqrt{1+\lambda}\tau\right) +
                    \frac{\sinh\left(\sqrt{1+\lambda}\tau\right)}
                    {\sqrt{1+\lambda}}\right]\, ,\quad\xi<-\tau\, .
                    \label{eq:Ctwo}
\end{equation}
For $\xi\in(-\tau,\tau)$, the situation is more complicated. First we 
write
\begin{eqnarray*}
\lefteqn{\cosh\left(\sqrt{p^2+\lambda}\tau\right) +
\frac{\sinh\left(\sqrt{p^2+\lambda}\tau\right)}{\sqrt{p^2+\lambda}}}\\
\quad
&=& \frac{1}{2}\left[\e^{\sqrt{p^2+\lambda}\tau}
    \left(1+\frac{1}{\sqrt{p^2+\lambda}}\right)+
    \e^{-\sqrt{p^2+\lambda}\tau}
    \left(1-\frac{1}{\sqrt{p^2+\lambda}}\right)\right]\, .
\end{eqnarray*}
For the first exponential, we can close the contour to the left and to 
the right for the second. Now, for both integrals, the integrant is not
an even function of $\sqrt{p^2+\lambda}$ and one shall get a
contribution of the cut. Integrating, one then eventually arrives to the
formula
\begin{eqnarray}
\hat\mu(\xi;\tau)
&=& 
\lambda\e^{-\tau}\left\{\e^{-\sqrt{1+\lambda}\tau}
    \left(1-\frac{1}{\sqrt{1+\lambda}}\right)\cosh(\xi)
    \rule{0mm}{10mm}\right. \label{eq:Cthree}\\
&+& \left.\rule{0mm}{10mm}2\int_0^{\sqrt{\lambda}}\frac{\d r}{\pi}
    \frac{\cos(r\xi)}{1+r^2}\left[
    \sinh\left(\sqrt{\lambda-r^2}\tau\right)+
    \frac{\cosh\left(\sqrt{\lambda-r^2}\tau\right)}
         {\sqrt{\lambda-r^2}}\right]\right\}\, ,\nonumber\\
\lefteqn{|\xi|<\tau\, .}\nonumber
\end{eqnarray}
Combining ~(\ref{eq:Cone},\ref{eq:Ctwo},\ref{eq:Cthree}) together
gives~(\ref{eq:muborne}).

}
\section*{Acknowledgements}
J.P. wishes to thank Dr. M. Droz for hospitality at the Department 
of Theoretical Physics of the University of Geneva where most of this research
has been done.

\vfill\eject
\begin{figure}

\end{figure}


\begin{thebibliography}{99}

\bibitem{redner}  S. Redner, to appear in {\it Proceedings of the 2nd 
International Colloquium on Quantum Field Theory and Stochastic 
Processes}, edt ., World Scientific (1996).

\bibitem{born} M. Droz, Some aspects of pattern formation 
in reaction-diffusion systems, in ``Diffusion Processes: Experiments, 
Theory, Simulations'', Lecture Notes in Physics {\bf 438}, 105-115, 
(1995), Springer-Velag.

\bibitem{wil}  D. Toussain and F. Wilczek, J. Chem. Phys {\bf 78},
2642, (1983)

\bibitem{racz} L. G\'alfi and Z. R\'acz, Phys.Rev {\bf A38}, 3151 (1988)

\bibitem{cordro} S. Cornell, M. Droz, and B. Chopard,  Phys. Rev.
{\bf A44}, 4826  (1991)


\bibitem{reddue} E. Ben-Naim, S. Redner and F. Leyvraz, Phys. Rev. Lett. 
{\bf 70}, 1890 (1993)


\bibitem{pia_uno} J. Piasecki,  Phys. Rev. {\bf E51}, 5535 (1995)

\bibitem{pia_due} M. Droz, P-A. Rey, L. Frachebourg and J. Piasecki,  
Phys. Rev. {\bf E51}, 5541 (1995)

\bibitem{pia_tre} M. Droz, P-A. Rey, L. Frachebourg and J. Piasecki,  
Phys. Rev. Lett. {\bf 75}, 160 (1995)

\bibitem{ef} Y. Elskens and H.L. Frisch, Phys. Rev. {\bf A31}, 3812 
(1985)


\bibitem{kopel} R. Kopelman, Sciences {\bf 241}, 1620 (1988)

\bibitem{buti} M. B\"uttiker and T. Christen, Phys. Rev. Lett. {\bf 75}, 
1895 (1995).

\bibitem{zrini} E. K\'arp\'ati-Smidr\'oczki, A. B\"uri and M. Zrinyi, 
Colloid. Polm. Sci. {\bf 273}, 857 (1995).


\bibitem{rus}A.P. Prudnikov, Yu.A. Brychkov and O.I. Marichev,
{\em Integrals and Series\/}, vol. 2 (Gordon and Breach Science 
Publishers, 1988), 311.

\end{thebibliography}
\end{document}